\documentclass[aps,twocolumn,nofootinbib,showpacs, prd]{revtex4}
\usepackage{graphicx}
\usepackage{amsfonts}
\usepackage{amssymb}
\usepackage{amsbsy}
\usepackage{amsmath}
\usepackage{latexsym}
\usepackage{natbib}
\usepackage{bm}
\usepackage{color}
\usepackage{float}

\begin{document}
\title{Phase Transition and Thermodynamic Stability in an Entropy-driven Universe}
\author{Soumya Chakrabarti\footnote{soumya.chakrabarti@vit.ac.in}}
\affiliation{School of Advanced Sciences \\
Vellore Institute of Technology \\ 
Tiruvalam Rd, Katpadi, Vellore, Tamil Nadu 632014 \\
India}

\pacs{}

\date{\today}

\begin{abstract}
Motivated by the notion that the mathematics of gravity can be reproduced from a statistical requirement of maximal entropy, we study the consequence of introducing an entropic source term in the Einstein-Hilbert action. For a spatially homogeneous cosmological system driven by this entropic source and enveloped by a time evolving apparent horizon, we formulate a modified version of the second law of thermodynamics. An explicit differential equation governing the internal entropy profile is found. Using a Hessian matrix analysis of the internal entropy we check the thermodynamic stability for a $\Lambda$CDM cosmology, a unified cosmic expanson and a non-singular ekpyrotic bounce. We find the mathematical condition for a second order phase transition during these evolutions from the divergence of specific heat at constant volume. The condition is purely kinematic and quadratic in nature, relating the deceleration parameter and the jerk parameter that chalks out an interesting curve on the parameter space. This condition is valid even without the entropic source term and may be a general property of any phase transition.
\end{abstract}

\maketitle

\section{Introduction}
A conversation between two relativists almost always involves a list of \textit{`usual suspects'}. The suspects are more-or-less associated with counter-intuitive phenomenologies related to gravity and it's unification with other fundamental interactions. General Theory of Relativity (GR) is used as the working language in these conversations. It provides a geometric portrayal of gravity starting simply from a requirement that the dynamical laws of our nature be generally covariant. In fact, it paves a way for Riemannian geometry in place of the standard Euclidean. The foundation principles were conceived about a century ago although the journey towards its status quo has been a continuous process ever since. Like any theory describing natural laws, GR also has scopes of improvement into a better version. For example, the questions regarding existence/formation of singularities \cite{collapse} and the behavior of geodesics at a length scale below the Planck scale are usually addressed using intuitive arguments, due to the lack of a resolute formalism of quantum gravity \cite{qg}. There are also loose ends remaining in most of the portrayals of a so-called cosmic expansion of the universe, related growth of masses and structures \cite{cosmo}. Finally, a unification of gravitational interaction with other fundamental forces is almost always left just as a mathematical exercise and deserves more attention. For instance, the now-detected Higgs boson, although celebrated mainly by particle physicists, can easily trigger follow-up investigations in the context of a theory of unification involving gravity. The heirarchy problem \cite{hierarchy} says that it is unphysical to accommodate the Higgs mass-scale within a standard model, embedded in Grand Unified Theory and this scenario echoes with the nature of cosmological constant problem. Intuitively, both of these entities are associated with the concept of vacuum energy density and should coexist in a common theoretical setup. \\

In this article, we briefly discuss the advantage of introducing an entropy-driven source term in GR. The equations of GR (or any fundamental interaction) are derived from a suitable action principle and an inference of thermodynamics can be drawn in as a phenomenological constraint. We adhere to a general imagination that the cosmic expansion is always enveloped within a time-evolving (apparent) horizon and explore it's thermodynamic properties. This notion has a similarity with black hole thermodynamics \cite{hawking, beken, bardeen, wald, paddy}, however, the idea of an evolving horizon seems to be a crucial difference. Our purpose is to find a generalized correlation between cosmic expansion and thermodynamic stability. which may have an additional purpose to approve/disapprove modifications of standard GR (for instance, models of dark energy, early universe or a unified expansion). An established approach is to study the \textit{Generalized Second Law of Thermodynamics (GSL)} and the nature of total entropy profile of the Universe \cite{thermo1, thermo2, thermo3}. During any local out-of-equilibrium phenomena, such as, particle production or formation of structures, an entropic force can easily modify these standard equations and alter the nature of total entropy content. We follow a more general motivation \cite{garcia1} and consider an omnipresent \textit{global} entropic force emerging from the degrees of freedom around the causal boundary of space [ref]. This formalism is capable of opening up new questions regarding the genesis of cosmic expansion and offer fresh intuitions on natural laws.  \\

Unless otherwise specified, we have used natural units for our calculations. We briefly review the basic equations of cosmology with a entropic source term in the next secion. We define the apparent horizon of a time evolving spherical, spatially homogeneous and isotropic geoemtry. Incorporating the Hayward-Kodama formalism \cite{hk1, hk2, hk3} we define the surface gravity and use its correlation with horizon temperature to calculate the total entropy content of the universe. Using a hessian matrix method \cite{anathermo1, duary} we further discuss the heat capacities, phase transitions and thermodynamic stability during a cosmic expansion. We discuss a few specific examples in section $3$ and subsequently conclude the article with some remarks, in section $4$.

\section{An Entropic Source Term in General Relativity}
To write a generally covariant setup unifying GR and non-equilibrium thermodynamics one should introduce a constraint in the Einstein-Hilbert action and generate a correction in the matter energy-momentum tensor. This correction comes in as an effective bulk viscosity term \cite{garcia1}. As a result the modified field equations accommodate a growth of entropy. As it has been checked in recent past, this growth can generate an effective negative pressure and drive the cosmic expansion without any unconventional matter component \cite{garcia2, garcia3}. A Lagrangian depending on generalized coordinates as well as the entropy $S$ should satisfy

\begin{equation}
\label{eq:var_prin}
\delta \int_{t_1}^{t_2} L(q, \dot{q}, S) dt = 0\,.
\end{equation}

The entropic force $F$ is implemented through variational constraint $\frac{\partial L}{\partial S} (q, \dot{q}, S) \delta S = \left<F(q, \dot{q}, S), \delta q \right>$, where the scalar product is written as $\left< \cdot , \cdot \right>$. The modified Euler-Lagrange equations alongwith the associated phenomenological constraint are written as \cite{garcia1}
\begin{equation}
\frac{d}{dt} \frac{\partial L}{\partial \dot{q}} - \frac{\partial L}{\partial q} = F (q, \dot{q}, S) ~,~ \frac{\partial L}{\partial S} \dot{S} = \left<F (q, \dot{q}, S), \dot{q} \right> \,.
\end{equation}

The constraints are necessary for this variational principle formulation to describe a \textit{closed} gravitational system, allowing an exchange of energy with the environment. It can be generalized further into systems allowing exchange of both energy and matter \cite{yoshimura}. For a spatially homogeneous case one can introduce the temperature of the system by defining $\frac{\partial L}{\partial S} = - T$. Assuming that the entropy function is homogeneous, we write the Einstein-Hilbert action as
\begin{equation}
\label{eq:grm}
\frac{1}{2\kappa} \int d^4x \sqrt{-g} R + \int d^4x \mathcal{L}_{m}(g_{\mu \nu}, S)\,.
\end{equation}

The stationary action principle leads to the following equations

\begin{eqnarray}\nonumber
&& \int d^4x \left[ \left(\frac{1}{2\kappa} \frac{\delta (\sqrt{-g}R)}{\delta g^{\mu \nu}} + \frac{ \delta \mathcal{L}_m}{\delta g^{\mu \nu}}\right)\delta g^{\mu \nu} + \frac{\partial \mathcal{L}_m}{\partial S} \delta S \right] = 0,\\&&
\frac{\partial L_{m}}{\partial S} \delta S = \frac{1}{2} F_{\mu \nu} \delta g^{\mu \nu}\,.
\end{eqnarray}
The second equation essentially works as the phenomenological constraint. The entropic force is written using a tensor $F_{\mu \nu}$ and a tensor density can be defined for the same as $F_{\mu \nu} = \int d^3x \sqrt{-g} f_{\mu \nu}$. Following usual variations one can derive the modified field equations as

\begin{equation}\label{eq:GREA}
G_{\mu\nu} = R_{\mu \nu} - \frac{1}{2} R g_{\mu \nu} = \kappa \left( T_{\mu \nu} -  f_{\mu \nu} \right).
\end{equation}

We interpret $f_{\mu\nu}$ as an effective bulk viscosity 
\begin{equation}
\label{eq:fmunu}
f_{\mu\nu} = \zeta\,D_\lambda u^\lambda \,(g_{\mu\nu}+u_\mu u_\nu) = \zeta\,\Theta\,h_{\mu\nu},
\end{equation}

and write the stress-energy tensor of the universe similar to an imperfect fluid
\begin{eqnarray}
\label{eq:Tmunu}
{\cal T}^{\mu\nu} &=& P\,g^{\mu\nu} + (\epsilon + P)u^\mu u^\nu -  \zeta\,\Theta\,h^{\mu\nu} \\ &=& \tilde P\,g^{\mu\nu} + (\epsilon + \tilde P)u^\mu u^\nu\,.
\end{eqnarray}

From the thermodynamic constraint, we can write the bulk viscosity coefficient as $\zeta = \frac{T}{\Theta}\frac{dS}{dV}$, where $\Theta$ is the expansion scalar and $V$ is the comoving volume. For a spatially homogeneous Friedmann-Robertson-Walker spacetime, $\zeta = T\dot S/(9H^2a^3)$, where $\Theta = 3H$ \cite{garcia2}. Note that, due to the non-trivial contribution of \textit{entropy per comoving volume}, components of the Einstein tensor (except $G_{00}$) now carry a growth in entropy. Spatial homogeneity indicates that there is no shear or vorticity. Therefore a family of geodesics converge/diverge obeying the modified Raychaudhuri equation as in
\begin{eqnarray}
\label{eq:Ray} \nonumber
\frac{D}{d\tau}\Theta + \frac{1}{3}\Theta^2 &=& - \sigma_{\mu\nu}\sigma^{\mu\nu} + \omega_{\mu\nu}\omega^{\mu\nu} - R_{\mu\nu} u^\mu u^\nu \\ &=& - \kappa\left(T_{\mu\nu}u^\mu u^\nu + \frac{1}{2} T^\lambda_{\ \,\lambda} - \frac{3}{2}\zeta\Theta\right) \\ &=& - \frac{\kappa}{2}(\epsilon + 3\tilde P) = -\frac{\kappa}{2}\left(\epsilon + 3P - 3T\frac{dS}{dV}\right) \nonumber\,.
\end{eqnarray}

Overall this formulation can also be found from the continuity equations. From the second law of thermodynamics, i.e., $T dS = d(\rho a^3) + p\,d(a^3) = 0$, an entropy-correction simply leads to 
\begin{equation}\label{eq:2LT}
\dot\rho + 3H(\rho+p) = \frac{T\dot S}{a^3}\,.
\end{equation}

Finally we write the modified FRW equations for a spatially flat metric as

\begin{eqnarray}\label{eq:FR1}
&&3\frac{\dot{a}}{a}^2 = \rho, \\&&
\frac{\ddot a}{a} = -\frac{4\pi G}{3}(\rho + 3p) + \frac{4\pi G}{3}\frac{T\dot S}{a^3H}.\label{eq:FR2}
\end{eqnarray}

\section{Phase Transition and Thermodynamic Stability}
We follow the approach of stability analysis of a model of expanding universe using specific heat capacities of the constituent elements \cite{thermo1, thermo2, thermo3}. This approach has been used extensively in recent times, to categorically rule out models of late-time acceleration which fail to produce a negative $C_V$ (specific heat at constant volume). In this anlysis, the thermodynamic stability of the universe is determined eventually using the total internal entropy and it's second order derivative \cite{anathermo1, anathermo2}. This is where, in the present work, the entropic correction in the field equations is expected to produce something different. We have found a second-order differential equation which governs the evolution of the total entropy of the universe (imagined as a time evolving sphere encircled by a horizon). Using a numerical solution of the same equation and suitable approximations, we further explore the allowed phase transitions of any constituent elements of the universe, using a hessian matrix method. We can take the horizon as an evolving null surface to allow the growth of degrees of freedom around the causal boundary. This null surface behaves like an apparent horizon defined using the condition $g^{\mu\nu} {Y}_{,\mu} {Y}_{,\nu} = 0$ ($Y$ is the radius of two-sphere). For a spatially flat and homogeneous metric we find that the condition leads to $Y = \frac{1}{H}$. We define the temperature of this horizon using a Hayward-Kodama \cite{hk1, hk2, hk3} formalism since the more popular formalism involving Hawking temperature is associated mainly with a static horizon and can allow no phase transitions during a cosmic expansion. Quite recently, using a Hayward-Kodama horizon formalism, the smooth transition from matter dominated deceleration into a late-time acceleration in $\Lambda$CDM cosmology has been correlated to a second order phase transition \cite{duary}. We find a more generalized set of equivalence in the presence of an entropy driven source term.  \\

A horizon can always be associated with a characteristic temperature through its surface gravity $\kappa$, following $T = \frac{\kappa}{2 \pi}$. In the Hayward-Kodama formalism, surface gravity $\kappa_{hk}$ for a spherically symmetric time evolving object depends on the area radius of two-sphere through the Kodama vector $K^a \equiv \epsilon^{ab} \nabla_b R$ \cite{kodama},
\begin{equation}\label{HK surgrav}
\frac{1}{2}g^{ab}K^c(\nabla_c K_a-\nabla_a K_c)= \kappa_{hk} K^b .
\end{equation}
$h^{ab}$ is the induced two-metric and $\epsilon^{ab}$ is its volume form. For a spatially flat geometry circumscribed by an evolving apparent horizon the surface gravity is written as $\kappa_{hk} = -\frac{1}{2H}(\dot{H}+2H^2)$ \cite{cai}. The apparent horizon therefore has a \textit{'Hayward-Kodama'} temperature \cite{hk1}
\begin{equation}\label{temp}
T = \frac{\mid{\kappa_{hk}\mid}}{2\pi} = \frac{2H^2+\dot{H}}{4\pi H}.
\end{equation}

The horizon entropy is proportional to surface area
\begin{equation}
S_h = 2\pi A ~,~ \dot{S_h}= -16 \pi^2 \frac{\dot{H}}{H^3}.
\end{equation}
To calculate the time derivative of entropy we have used the area of apparent horizon $A = 4\pi Y^2$. We define $S_{\mbox{in}}$ and $U$ to write the entropy and total internal energy of the fluid inside the horizon. If the system is not necessarily adiabatic, the first principle generates a thermodynamic constraint $TdS_{\mbox{in}} = dU + pdV + Vdp$. $V$ is the volume of the fluid enclosed by the evolving apparent horizon $V =\frac{4}{3}\pi Y^3 =\frac{4}{3}\pi \frac{1}{H^3}$. Altogether, we can calculate the first order change of entropy with cosmic time as

\begin{equation}\label{ent}
\dot{S}_{\mbox{in}} = \frac{1}{T_h} \left[(\rho+p)\dot{V} + (\dot{\rho}+\dot{p})V\right].
\end{equation}

Using the Hayward-Kodama temperature as in Eq. (\ref{temp}) and entropy-modified field Eqs. (\ref{eq:FR1}), (\ref{eq:FR2}), we derive a second order differential equation governing ${S}_{\mbox{in}}$.
\begin{eqnarray}\label{ent-rate1}\nonumber
&&\frac{d^{2}{S}_{\mbox{in}}}{dt^2} - \frac{d{S}_{\mbox{in}}}{dt} \Bigg[ \frac{9Ha^3}{4\pi} + 3Ha^3 \Bigg\lbrace \frac{\dot{H}}{a^{3}H^5} - \frac{\dot{T_h}}{3T_{h}a^{3}H} \\&&
 + \frac{1}{a^3} + \frac{\dot{H}}{3a^{3}H^{2}}\Bigg\rbrace \Bigg] + \frac{12\pi H^2 a^3}{(2H^2 + \dot{H})}\left( 6\frac{\dot{H}^2}{H^4} - 2\ddot{H}\right) = 0.
\end{eqnarray}

\begin{figure}[t!]
\begin{center}
\includegraphics[angle=0, width=0.40\textwidth]{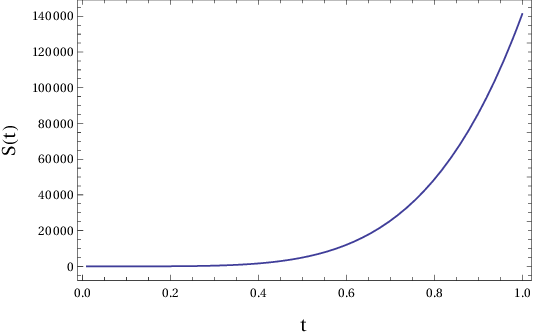}
\includegraphics[angle=0, width=0.40\textwidth]{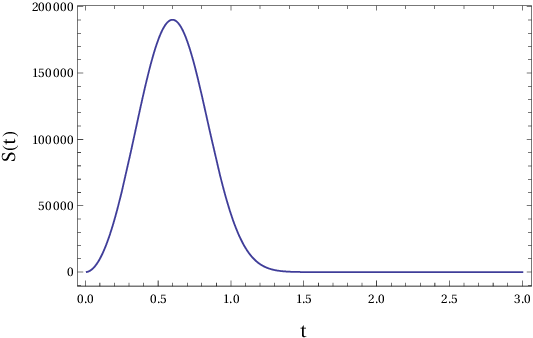}
\includegraphics[angle=0, width=0.40\textwidth]{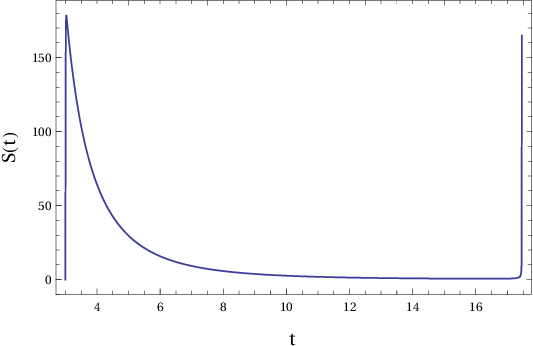}
\caption{Evolution of entropy ($S$) as a function of cosmic time for (i) top : an ever-accelerating model ($a(t) \sim t^{\frac{4}{3}}$), (ii) middle : decelerating model ($a(t) \sim t^{\frac{2}{3}}$) and (iii) bottom : a unified expansion model ($a(t) \sim \exp \Big[H_{0}t -\frac{H_{1}}{(n-1)t^{n-1}}\Big]$).}
\label{ent_1}
\end{center}
\end{figure}

A direct numerical solution of the above equation leads to the plots in Fig. \ref{ent_1}. The numerical solution depends on the functional form of $a(t)$, i.e., the particular model of acceleration ine chooses to work with. The graph on top shows the entropy profile for an ever-accelerating universe ($a(t) \sim t^{\frac{4}{3}}$). The graph in middle is for a forever decelerating universe ($a(t) \sim t^{\frac{2}{3}}$). The graph in bottom is for a model suggestive of unifies cosmic time history, an early acceleration followed by a matter dominated deceleration and finally, a late-time acceleration. The entropy profile is always positive and regular, unless the scale factor tends to zero or infinity (around big bang or a future big crunch singularity) It also appears that the entropy (can also be imagined as a scalar field) increases rapidly whenever the universe is driven into a phase of acceleration. An epoch of deceleration is always accompanied by a decay in entropy. One can see this happening in the unified picture of cosmic expansion as well (the numerical solution is for a scale factor $a(t) \sim \exp \Big[H_{0}t -\frac{H_{1}}{(n-1)t^{n-1}}\Big]$). \\

A study of thermodynamic stability involves maximization of entropy of the fluid enclosed by the horizon. Deriving the principle minors of a Hessian matrix of entropy \cite{anathermo2}, the mathematical requirements for this maximization are written as the following inequalities
\begin{eqnarray} 
&& (i)\frac{\partial^2{S_{\mbox{in}}}}{\partial U^2} \equiv = -\frac{1}{T^2C_V} \leq 0, \label{condi1}\\&&\nonumber
(ii)\frac{\partial^2{S_{\mbox{in}}}}{\partial U^2}\frac{\partial^2{S_{\mbox{in}}}}{\partial V^2}- \left(\frac{\partial^2{S_{\mbox{in}}}}{\partial U\partial V}\right)^2 \\&&
\equiv \frac{1}{C_VT^3V\beta_T} \equiv \alpha \geq 0. \label{condi2}
\end{eqnarray}

We also need to define
\begin{equation}\label{cv}
C_V = T\left(\frac{\partial S_{\mbox{in}}}{\partial T}\right)_V ;~ C_P = T\left(\frac{\partial S_{\mbox{in}}}{\partial T}\right)_P ;~ \beta_T = -\frac{1}{V}\left(\frac{\partial V }{\partial P}\right)_T,
\end{equation}

where $C_V$ and $C_P$ are the heat capacities at constant volume and constant pressure. $\beta_T$ is the isothermal compressibility. To explore the profile of specific heat capacities analytically, we study the differential Eq. (\ref{ent-rate1}) around the extremas of entropy, i.e., points where $\frac{d^{2}{S}_{\mbox{in}}}{dt^2} \sim 0$. In such a case the calculations become simpler as we can find an expression for the first order change in entropy,

\begin{equation}\label{sdot}
\dot{S}_{\mbox{in}} = \frac{\frac{12\pi H^{2}a^3}{(2H^2 + \dot{H})}\left(6\frac{\dot{H^2}}{H^4} - 2\ddot{H}\right)}{\left[\frac{9Ha^3}{4\pi} +3Ha^3 \left\lbrace \frac{\dot{H}}{a^3 H^5} -\frac{\dot{T}}{3Ta^3 H} + \frac{\dot{H}}{a^3 H^2} + \frac{1}{a^3} \right\rbrace\right]}.
\end{equation}

Using Eqs. (\ref{ent}) and (\ref{sdot}) we calculate 
\begin{eqnarray}\label{cv-expr}
&& C_V = V\left(\frac{\partial \rho}{\partial T}\right)_V + V\left(\frac{\partial p}{\partial T}\right)_V \nonumber \\&&
= \frac{16\pi^2}{3H}\frac{\left\lbrace -2\ddot{H} +\dot{S}\left(\frac{\dot{T}}{3a^3 H} - \frac{T}{a^3} - \frac{T\dot{H}}{3a^3 H^2}\right)\right\rbrace}{2H^2\dot{H}+H\ddot{H}-\dot{H}^2}.
\end{eqnarray}
 
\begin{eqnarray}\label{cp-expr}
&& C_P =  V\left(\frac{\partial \rho}{\partial T}\right)_P+(\rho+P)\left(\frac{\partial V}{\partial T}\right)_P \nonumber \\&&
= \frac{32\pi^2 \dot{H} \left\lbrace 1 +\frac{1}{2H^2} \left(2\dot{H} - \frac{T\dot{S}}{3a^{3}H}\right)\right\rbrace}{(2H^2\dot{H}+H\ddot{H}-\dot{H}^2)}.
\end{eqnarray}
 
\begin{eqnarray}\label{beta-expr}
&& \beta_T = -\frac{1}{V}\left(\frac{\partial V }{\partial P}\right)_T \nonumber\\&&
= \frac{\big(-\frac{3\dot{H}}{H}\big)}{2\ddot{H} + 6\dot{H}H + \frac{3Ha^3 \Big(6\frac{\dot{H}^2}{H^4} - 2\ddot{H}\big)\big(\frac{1}{a^3} + \frac{\dot{H}}{2a^3 H^2}\big)}{\big[\frac{9Ha^3}{4\pi} + 3Ha^3 \big(\frac{\dot{H}}{a^3 H^5} + \frac{\dot{H}}{a^3 H^2} + \frac{1}{a^3}\big)\big]}}.
 \end{eqnarray}

\begin{figure}[t!]
\begin{center}
\includegraphics[angle=0, width=0.40\textwidth]{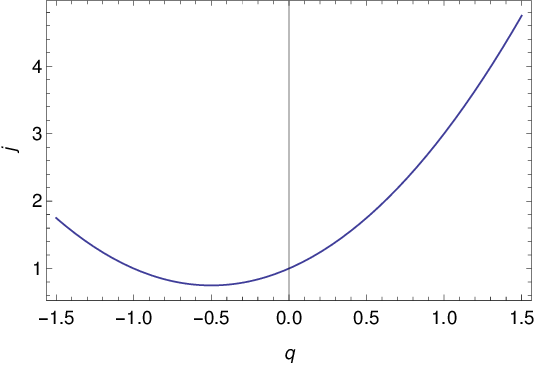}
\includegraphics[angle=0, width=0.40\textwidth]{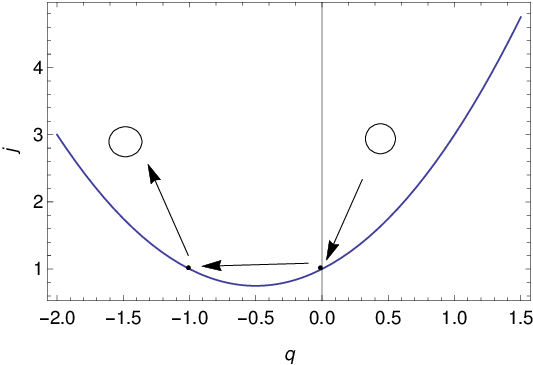}
\caption{Plot of Eq. (\ref{dimlesscondi}) : the kinematic condition (jerk vs deceleration) of phase transition for the universe.}
\label{jandq}
\end{center}
\end{figure}
 
The expression in Eqs. (\ref{cv-expr}) and (\ref{cp-expr}) can be written more explicitly by writing temperature and entropy gradient as a function of $H$, using Eqs. (\ref{temp}) and (\ref{sdot}). The bottomline is that the heat capacities and the thermodynamic stability (as in Eqs. (\ref{condi1}) and (\ref{condi2})) depend on the scale factor describing the universe. The primarly motivation is to investigate any divergence of $C_V$ and/or $C_P$. Such a divergence implies a second order phase transition and clearly projects an underlying thermodynamic nature of gravity. Looking at Eq. (\ref{cv-expr}), we note that a divergence can only happen at points where
\begin{equation}\label{condiii}
\ddot{H}H + 2H^{2}\dot{H} - \dot{H}^2 = 0.
\end{equation}
We use the definitions of dimensionless kinematic parameters deceleration and jerk
\begin{equation}
q = -\frac{\big(\frac{d^{2}a}{dt^2}\big)a}{{\big(\frac{da}{dt}\big)}^2} ~,~ j = \frac{\big(\frac{d^{3}a}{dt^3}\big)a^2}{{\big(\frac{da}{dt}\big)}^3},
\end{equation}
and write Eq. (\ref{condiii}) purely as a kinematic condition relating the parameters
\begin{equation}\label{dimlesscondi}
q^2 + q + 1 - j = 0.
\end{equation}

Eq. (\ref{dimlesscondi}) is remarkable in the sense that it defines a simple quadratic relation in the parameter space describing an expansion of the universe. The notion that such a relation defines the critical condition of any second order phase transition during the evolution of the universe, is one of a kind. It remains to be seen if this kinematic condition accompanies any kind of phase transition, even beyond the scope of gravity. Intuitively, can have a more fundamental root encoded within the kinematics of a flow/geodesic motion, same as the Raychaudhuri equation \cite{rc, sc}. An interesting feature is found from the above condition: if one puts the jerk parameter equal to $1$, there can be two values of $q$ for which a phase transition is realized, $q = 0$ and $q = -1$. One may recall that $j = 1$ gives an exact $\Lambda CDM$ model. For a $\Lambda$CDM model in particular, $q = 0$ denotes the point of deceleration-into-acceleration transition of the universe. The possibility of a second order phase transition at this point was very recently discussed by Duary, Banerjee and Dasgupta \cite{duary}. The $q = -1$ point raises more curiosity. We plot Eq. (\ref{dimlesscondi}) as a jerk vs deceleration curve in Fig. \ref{jandq}. For any particular value of jerk, the two values of deceleration parameter allowing a phase transition can be marked with dots (in Fig. \ref{jandq} it is done for $\Lambda$CDM). We imagine and compare the tendancy of our universe to evolve towards a phase transition with a ball, being dropped on the parametric curve of Fig. \ref{jandq}. Assuming all motions to be perfectly elastic, as long as the ball does not touch the track/curve, there is no phase transition, i.e., the universe remains in the same phase. The moment it hits the curve, a phase transition can happen. The ball is reflected from the point of first impact (right quadrant) and once again hits the track at a second impact point (left quadrant) beforing ejection. The exact points of contact depends on the time/redshift values at which the kinematic parameters start obeying Eq. (\ref{dimlesscondi}), i.e., depends on the scale factor.  \\

The conditions in Eqs. (\ref{condiii}) and (\ref{dimlesscondi}) are easily reproduced for standard cosmological scenario as well. The readers may cross-check using the recent result in \cite{duary}, that a divergence of $C_V$ is always found at the zero-s of $2H^2\dot{H}+H\ddot{H}-\dot{H}^2$. Depending on the theory of gravity, we can also write this condition using the components of energy-momentum tensor. For standard GR, using the Raychaudhuri equation,
\begin{equation}
3\frac{\ddot{a}}{a} = -\frac{(\rho+ 3p)}{2},
\end{equation} 
we may write Eq. (\ref{condiii}) as
\begin{equation}\label{condiiem}
\frac{\sqrt{\rho}}{3\sqrt{3}}\frac{d}{dt}(\rho + 3p) + (\rho + p)^2 = 0.
\end{equation}
This equation can be solved for different choices of equation of state and accordingly, the density at which a phase transition should occur, can be determined. For instance, for a perfect fluid given by $p = \omega \rho$, it is straightforward to solve Eq. (\ref{condiiem}) and find that the phase transition density is given by
\begin{equation}
\rho_{pt} \sim \frac{(1+3\omega)^2}{27(1+\omega)^4 (t - t_0)^2}.
\end{equation}
Similarly, for a universe with entropy-driven acceleration, the condition for phase translation can be re-written using Eq. (\ref{eq:FR2}) as
\begin{equation}\label{condiiem1}
\frac{\sqrt{\rho}}{3\sqrt{3}} \frac{d}{dt}\left[\rho + 3p + 9\sqrt{3}\sqrt{\rho}\zeta \right] + (\rho + p + 3\sqrt{3} \sqrt{\rho} \zeta)^2 = 0.
\end{equation}
The readers might recall that apart from standard pressure and energy density, there is an entropy-driven bulk viscosity correction in the energy-momentum tensor and naturally it contributes in the critical condition.
 
\begin{figure}[t!]
\begin{center}
\includegraphics[angle=0, width=0.40\textwidth]{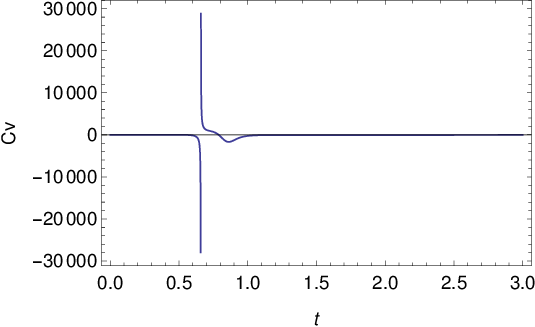}
\includegraphics[angle=0, width=0.40\textwidth]{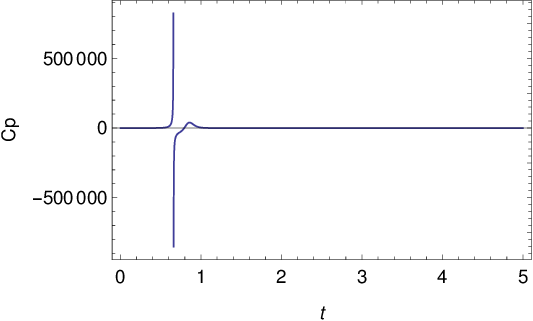}
\includegraphics[angle=0, width=0.40\textwidth]{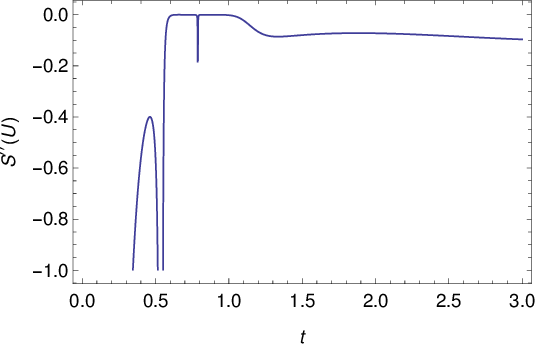}
\includegraphics[angle=0, width=0.40\textwidth]{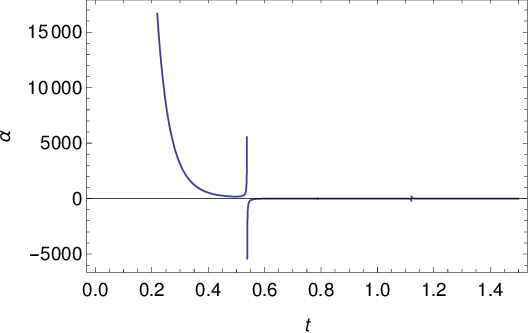}
\caption{Plots of $C_V$, $C_P$, $\frac{\partial^2{S}}{\partial U^2}$ and $\alpha$ as a function of cosmic time for a $\Lambda$CDM cosmology ($a(t) = \left(\frac{\Omega_m}{\Omega_{vac}}\right)^{1/3}\sinh^{2/3}(t/t_0)$).}
\label{LCDM}
\end{center}
\end{figure}

\section{Specific Examples}
As a continuation, in this section we explore the evolutions of $C_V$, $C_P$, $\frac{\partial^2{S}}{\partial U^2}$ and $\alpha$, for three qualitatively different scenarios of cosmic evolution.

\subsection{$\Lambda$CDM}
A $\Lambda$CDM model exhibits a late-time accelerated expansion preceded by a matter dominated deceleration. It is by far the most favored model for the present accelerating universe and fits well with astrophysical evidences such as Hubble free luminosity measurement \cite{cosmo}. It is well described by the scale factor 

\begin{equation}
a(t) = \left(\frac{\Omega_m}{\Omega_{vac}}\right)^{1/3}\sinh^{2/3}(t/t_0).
\end{equation}

In a matter-dominated era the above scale factor behaves like a power law expansion $\propto t^{2/3}$ and during late-times mimics an exponential evolution $\propto \exp(Ht)$. The Hubble function behaves as $H(t) = \frac{2}{3}\coth (t/t_0)$. Using the expression of scale factor and Hubble function in Eqs. (\ref{cv-expr}), (\ref{cp-expr}), (\ref{beta-expr}), (\ref{condi1}) and (\ref{condi2}), we plot the $C_V$, $C_P$, $\frac{\partial^2{S}}{\partial U^2}$ and $\alpha$ in Fig. \ref{LCDM}. The discontinuity in $C_V$ can be clearly seen around a point where the scale factor starts describing a late-time acceleration. We are not converting the expressions into functions of redshift but that can easily be done. The discontinutiy of $C_V$ means a second order phase transition and it coincides with the transiton of universe from deceleration into acceleration (For $\Lambda$CDM, at $j = 1$ and $q = 0$). $C_P$ is negative during late-times. More importantly, we find that the entropy correction to cosmological equations (and to $C_V$ as in Eq. (\ref{cv-expr})) makes this model thermodynamically stable, as both of the inequalities $\frac{\partial^2{S}}{\partial U^2} \leq 0$ and $\alpha \geq 0$ are satisfied, atleast during late-times (contrary to the standard cosmological case where these are never satisfied). \\

\subsection{A Unified Cosmic Expansion}

\begin{figure}[t!]
\begin{center}
\includegraphics[angle=0, width=0.40\textwidth]{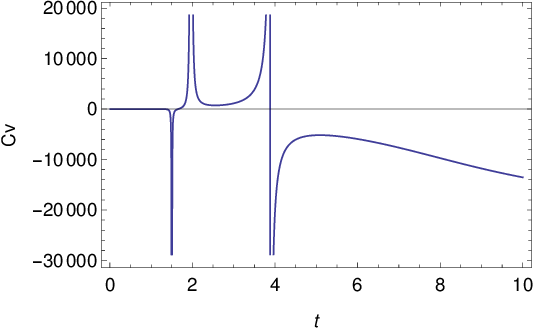}
\includegraphics[angle=0, width=0.40\textwidth]{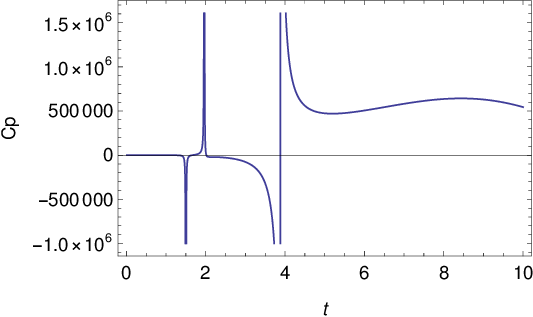}
\includegraphics[angle=0, width=0.40\textwidth]{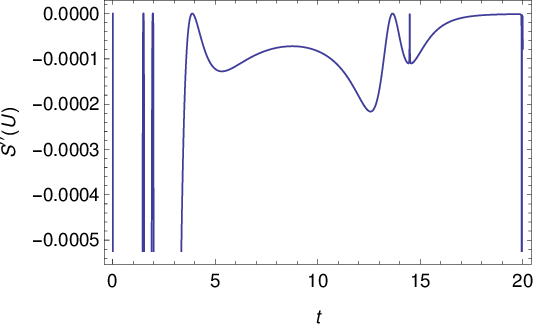}
\includegraphics[angle=0, width=0.40\textwidth]{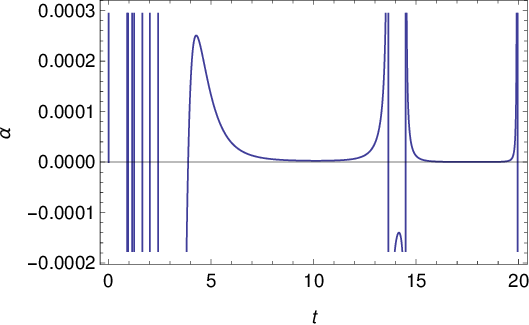}
\caption{Plots of $C_V$, $C_P$, $\frac{\partial^2{S}}{\partial U^2}$ and $\alpha$ for a unified description of cosmic expansion ($a(t) = a_{0}\exp \Big[H_{0}t -\frac{H_{1}}{(n-1)t^{n-1}}\Big]$).}
\label{unified}
\end{center}
\end{figure}

A model of unified cosmic expansion that can smoothly toggle between deceleration/acceleration at different time values is indeed interesting. It is perhaps the ultimate goal of cosmology to identify such a model fit it with the available data sets of astrophysical observation. Although a perfect functional form for this evolution is yet to be found, we proceed with the discussion using a simple toy model 

\begin{eqnarray}\label{1.13}
&& H(t) = H_{0}+\frac{H_{1}}{t^{n}} \\&&
a(t) = a_{0}\exp \Big[H_{0}t -\frac{H_{1}}{(n-1)t^{n-1}}\Big].
\end{eqnarray}

The model parameters $H_{0}$, $H_{1}$, $n$ and $a_{0}$ can change the overall scale of the evolution but mostly, do not affect the qualitative nature. $t \sim 0$ marks an event alike big-bang where $a(t) \sim 0$. For parameter values $n = 4$, $H_{0} = 1$, $H_{1} = 0.05$ and $a_{0} = 1$, a rapid inflation is realized until $t \sim 0.05$. Phases of deceleration and the present acceleration should follow and they are understood from an evolving equation of state, written as
\begin{equation}\label{1.14}
w_{\rm eff} = -1-\frac{2\dot{H}}{{3H}^{2}} = -1+\frac{2nH_{1}t^{n-1}}{\big(H_{0}t^{n}+H_{1}\big)^{2}}.
\end{equation}

$w_{\rm eff} \Rightarrow -1$ for $t \Rightarrow 0$ as well as $t \Rightarrow \infty$. These can mimic the phase of an early and a late-time acceleration. Two critical points where the nature of evolution changes from acceleration to deceleration or vice versa are found from the zero-s of 
\begin{equation}
\frac{\ddot a}{a} = \dot H +H^{2}=-\frac{nH_{1}}{t^{n+1}} + \left( H_{0}+\frac{H_{1}}{t^{n}}\right) ^{2}\ . \label{1.15}
\end{equation}

The zero-s are found at
\begin{equation}
t_{\pm} \approx \left[ \sqrt{nH_{1}} \,\, \frac{\left( 1\pm
\sqrt{1-\frac{4H_{0} }{n}} \, \right)}{2H_{0}} \, \right]^{2/n},
\label{1.16}
\end{equation}

where a condition of $\frac{4H_{0}}{n} \leq 1$ is imposed for obvious reasons. From the two time values of critical points, one can portray a unified time history : the phase $0 < t < t_{-}$ defines an early inflation, the phase $t_{-} < t < t_{+}$ defines a matter-dominated deceleration and the present acceleration is realized for all $t > t_{+}$. Using Eq. (\ref{1.13}), we plot the $C_V$, $C_P$, $\frac{\partial^2{S}}{\partial U^2}$ and $\alpha$ in Fig. \ref{unified}. There are two discontinuities in $C_V$ around the time values where the universe moves from an accelerated into a decelerated phase or vice-versa. $C_P$ is positive during comsic acceleration and negative during the preceding deceleration. This model is thermodynamically stable since $\frac{\partial^2{S}}{\partial U^2}$ is always negative. However, there are non-trivial disontonuities in the evolution of $\frac{\partial^2{S}}{\partial U^2}$ and $\alpha$. Crucially, the phase transitions for this case do not coincide with $q = 0$, since the evolution is not exactly $\Lambda$CDM.

\subsection{An Ekpyrotic Bounce}
We explore another alternative notion where the universe does not necessarily spring out of a \textit{big bang} singularity, but rather gets driven into the initial expansion through a non-singular bounce. For all $n < \frac{1}{6}$, a scale factor $\left[1+a_0t^2 \right]^n$ can describe a bounce ($t < 0$ domain is for the preceding contraction) followed by an expansion. The minima of this scale factor is realized at the bounce instant ($t = 0$). In order to unify this with a dark energy driven acceleration at late-times, one can include an exponential function $\sim \exp\left( \frac{1}{\beta-1}(t_s-t)^{1-\beta} \right)$. The two of them can be unified using a scale factor \cite{ekpyrotic}
\begin{equation}
a(t) = \left[ 1+a_0t^2 \right]^n\exp\left( \frac{1}{\beta-1}(t_s-t)^{1-\beta} \right). \label{eq7}  
\end{equation}
The Hubble function is also easily derived as
\begin{equation}
H(t)=\frac{2a_0nt}{1+a_0t^2}+\frac{1}{(t_s-t)^\beta} \label{eq8}
\end{equation}

\begin{figure}[t!]
\begin{center}
\includegraphics[angle=0, width=0.40\textwidth]{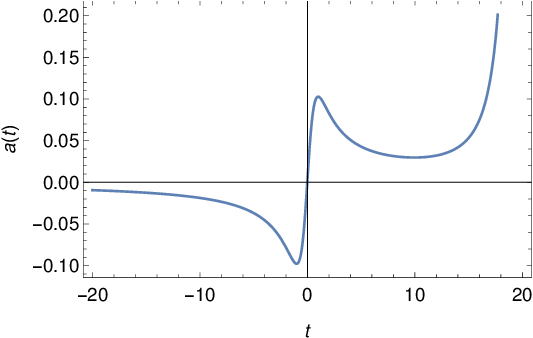}
\caption{Scale factor for an universe exhibitng ekpyrotic bounce, $a(t) = \left[ 1+a_0t^2 \right]^n\exp\left( \frac{1}{\beta-1}(t_s-t)^{1-\beta} \right)$.}
\label{scale_ekpy}
\end{center}
\end{figure}

\begin{figure}[t!]
\begin{center}
\includegraphics[angle=0, width=0.40\textwidth]{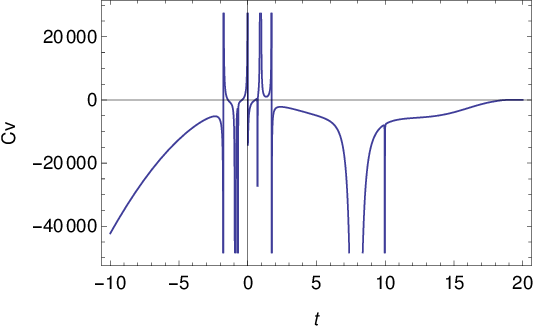}
\includegraphics[angle=0, width=0.40\textwidth]{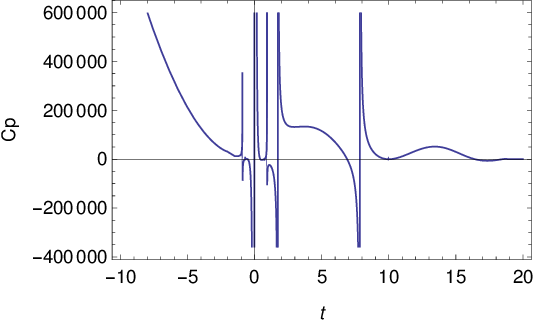}
\includegraphics[angle=0, width=0.40\textwidth]{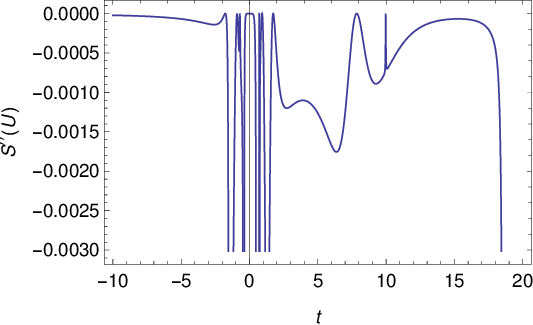}
\includegraphics[angle=0, width=0.40\textwidth]{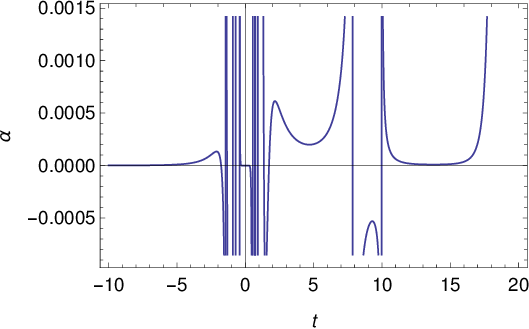}
\caption{Plots of $C_V$, $C_P$, $\frac{\partial^2{S}}{\partial U^2}$ and $\alpha$ for an universe exhibitng ekpyrotic bounce ($a(t) = \left[ 1+a_0t^2 \right]^n\exp\left( \frac{1}{\beta-1}(t_s-t)^{1-\beta} \right)$).}
\label{ekpy}
\end{center}
\end{figure}

The pattern of the scale factor is drawn in Fig. \ref{scale_ekpy} for convenience. Using Eq. (\ref{eq7}) and (\ref{eq8}) we plot the $C_V$, $C_P$, $\frac{\partial^2{S}}{\partial U^2}$ and $\alpha$ in Fig. \ref{ekpy}. There are multiple discontinuities in $C_V$ around the time value where the universe sees a transition from the preceding collapsing phase into an expanding phase. We can correlate this with a domain infested with rapid phase transitions. Apart from the regions where there are discontinuities, $C_P$ remains positive. The model shows signs of thermodynamic stability since $\frac{\partial^2{S}}{\partial U^2}$ is always negative. However, due to the large number of discontinuities in $\alpha$, the acceleraton part of a unified ekpyrotic bounce remains questionable. Using a better form of scale factor that can reciprocate with a similar qualitative behavior, one may be able to address this issue.

\section{Conclusion}
This article is motivated by the fact that some laws of General Theory of Relativity, in particular, laws related to the behavior of black holes are very much suggestive of thermodynamic systems. This comparison finds additional boost in quantum field theoretic calculations in Rindler spacetime. It provokes a thought that gravity is perhaps better described not just as a fundamental force, but as a macroscopic, thermodynamic phenomenon on large scales. Atleast, the mathematics of the theory should be reproducible from a statistical requirement of maximal entropy. This notion can be used to explain the absence of an equivalence principle for charges defined in other fundamental forces except for an inertial/gravitational mass. It can also provide a fair alibi for the leverage gravity has on cosmological scales inspite of being subdued compared to the short range fundamental forces. The prevalent issues related to a renormalisable quantum field theories and the shortcomings of a quantum gravity can be also be addressed using this notion. Although questions related to the lack of solutions involving formation of structure, gravitational collapse or a space-time singularity may remain, this much is motivating enough that gravity, naturally associated with a classical scale, can be portrayed as an emergent statistical aggregation of multi-particle systems. \\

Most of the discussions in this article involve standard curiosities regarding the total internal entropy and thermodynamic stability of a general relativistic model of cosmic expansion. However, the theoretical setup has a crucial modification. It incorporates a thermodynamic source term in the Einstein-Hilbert action and as a consequence, an entropic correction leads to a modified set of field equations. Expecting that the cosmic evolution should be enveloped by a time-evolving null surface, we employ the Kodama vector to define the surface gravity and correlate the same with entropy. Formulating a modified version of second law of thermodynamics (for a system that is not necessarily adiabatic) we write explicitly the differential equation governing the total internal entropy. The primary motivation has been to check if different phases of an expanding universe can exhibit thermodynamic stability. The stability depends on different forms of the scale factor, however, it is found that due to the presence of the entropic source term, the models can be thermodynamically stable. Mathematically, this is realized by constructing a Hessian matrix of total internal entropy and checking if $\frac{\partial^2{S_{\mbox{in}}}}{\partial U^2} \leq 0$. We have checked three qualitatively different cases as examples, (i) a $\Lambda$CDM cosmology, (ii) a model that can unify (even if just as a toy model) the three epochs of cosmic expanson with an acceleration-deceleration-acceleration order and (iii) an ekpyrotic bounce where there is no big bang singularity and the universe gets driven into an early inflation by a bounce from the preceding collapsing phase. \\

The most curious outcome is that any phase transition of the expanding universe is not necessarily realized at a single point, but on a curve defined in the parameter space. The condition is written as Eq. (\ref{dimlesscondi}) which is remarkably, just a kinematic condition relating the deceleration and the jerk parameter. It is valid in general, i.e., an exactly similar equation can be found even without the entropic source term. It further suggests that for any solution of the scale factor, phase transitions should happen at two values of the deceleration parameter (since the equation is quadratic in $q$). The parameter space plot of this condition helps portray an interesting analogy. The tendancy of the universe towards a phase transition can be compared to a ball being dropped and having a perfectly elastic collision on a track like the one shown in Fig. \ref{jandq}. After the ball is dropped, it can hit the curve atmost twice before jumping up again. Precisely, apart from these two points no other parameter values allow the universe to satisfy a condition for phase transition. The exact location of the points in the parameter space depends on the solution we are working with. Our intuition is that perhaps this condition is not just a consequence of the field equations of gravity but rather a property of any phase transition. We shall try to draw such an inference from a simple thermodyanmic point of view, in a subsequent work. The condition can also relate the components of an effective energy momentum tensor having energy density, pressure and bulk viscosity as components. Using the condition of geodesic deviation we derive this correlation to be
\begin{equation}
\frac{\sqrt{\rho}}{3\sqrt{3}} \frac{d}{dt}\left[\rho + 3p + 9\sqrt{3}\sqrt{\rho}\zeta \right] + (\rho + p + 3\sqrt{3} \sqrt{\rho} \zeta)^2 = 0.
\end{equation}

This can work as a crucial constraint while delving into further studies of a modified gravity theories with entropic source term. For instance, the description of a gravitational collapse and subsequent formation of singularities/exotic objects might be an avenue where one might attempt to see the effect of this constraint. Since a collapse can naturally lead a general relativistic sphere towards Planck length scale, one might use this motivation to think about thermodynamic constraints in a formalism of quantum gravity \cite{sgcsc} as well. \\

\section*{Acknowledgement}
The author acknowledges Vellore Institute of Technology for the financial support through its Seed Grant (No. SG20230027), year $2023$.

\end{document}